\documentclass[twocolumn,tighten,times]{aastex62}
\usepackage{mathrsfs}
\usepackage{amsmath}
\usepackage{url}
\usepackage[normalem]{ulem}
\usepackage{graphicx}
\usepackage{float}

\usepackage{natbib}
\usepackage{color}
\bibpunct{(}{)}{;}{a}{}{,}

\newcommand\aproxgt{\mathrel{%
      \rlap{\raise 0.511ex \hbox{$>$}}{\lower 0.511ex \hbox{$\sim$}}}}
\newcommand\aproxlt{\mathrel{%
      \rlap{\raise 0.511ex \hbox{$<$}}{\lower 0.511ex \hbox{$\sim$}}}}

\newcommand{\ignore}[1]{}
\newcommand{\storm}{{STORM}}

\newcommand{\ngc}{NGC\,5548}
\newcommand{\et}{{et~al.\, }}

\newcommand{\lya}{\mbox{\rm Ly$\alpha$}}

\newcommand{\cii}{\mbox{\rm C\,{\sc ii}}}
\newcommand{\ciii}{\mbox{\rm C\,{\sc iii}}}
\newcommand{\civ}{\mbox{\rm C\,{\sc iv}}}
\newcommand{\siII}{\mbox{\rm Si\,{\sc ii}}}
\newcommand{\siIII}{\mbox{\rm Si\,{\sc iii}}}
\newcommand{\siIV}{\mbox{\rm Si\,{\sc iv}}}

\newcommand{\nv}{\mbox{\rm N\,{\sc v}}}

\shorttitle{AGN \storm\ ~X. Absorption Line Holiday in \ngc}
\shortauthors{Dehghanian \et}

\begin{document}

\title{Space Telescope and Optical Reverberation Mapping Project. X. Understanding the 
Absorption-Line Holiday in NGC 5548}

\author{M.~Dehghanian}
\affiliation{\ignore{UKy}Department of Physics and Astronomy, The University of Kentucky, Lexington, KY 40506, USA}
\author{G.~J.~Ferland}
\affiliation{\ignore{UKy}Department of Physics and Astronomy, The University of Kentucky, Lexington, KY 40506, USA}
\author{G.~A.~Kriss}
\affiliation{Space Telescope Science Institute, 3700 San Martin Drive, Baltimore, MD 21218, USA}
\author{B.~M.~Peterson}
\affiliation{Space Telescope Science Institute, 3700 San Martin Drive, Baltimore, MD 21218, USA}
\affiliation{Department of Astronomy, The Ohio State University, 140 W 18th Ave, Columbus, OH 43210, USA}
\affiliation{Center for Cosmology and AstroParticle Physics, The Ohio State University, 191 West Woodruff Ave, Columbus, OH 43210, USA}
\author{S.~Mathur}
\affiliation{Department of Astronomy, The Ohio State University, 140 W 18th Ave, Columbus, OH 43210, USA}
\affiliation{Center for Cosmology and AstroParticle Physics, The Ohio State University, 191 West Woodruff Ave, Columbus, OH 43210, USA}
\author{M.~Mehdipour}
\affiliation{\ignore{SRON}SRON Netherlands Institute for Space Research, Sorbonnelaan 2, 3584, CA Utrecht, The Netherlands}
\author{F.~Guzm\'{a}n}
\affiliation{\ignore{UKy}Department of Physics and Astronomy, The University of Kentucky, Lexington, KY 40506, USA}
\author{M.~Chatzikos}
\affiliation{\ignore{UKy}Department of Physics and Astronomy, The University of Kentucky, Lexington, KY 40506, USA}
\author{P.~A.~M.~van~Hoof}
\affiliation{\ignore{Belgium}Royal Observatory of Belgium, Ringlaan 3, B-1180 Brussels, Belgium}
\author{R.~J.~R.~Williams}
\affiliation{\ignore{Williams} AWE plc, Aldermaston, Reading RG7 4PR, UK}

\author{N.~Arav}
\affiliation{Department of Physics, Virginia Tech, Blacksburg, VA 24061, USA}

\author{A.~J.~Barth}
\affiliation{\ignore{UCI}Department of Physics and Astronomy, 4129 Frederick Reines Hall, University of California, Irvine, CA 92697, USA}

\author{M.~C.~Bentz}
\affiliation{\ignore{Georgia}Department of Physics and Astronomy, Georgia State University, 25 Park Place, Suite 605, Atlanta, GA 30303, USA}

\author{S.~Bisogni}
\affiliation{Department of Astronomy, The Ohio State University, 140 W 18th Ave, Columbus, OH 43210, USA}
\affiliation{\ignore{Arcetri}Osservatorio Astrofisico di Arcetri, largo E. Fermi 5, 50125, Firenze, Italy}
\affiliation{\ignore{CfA}Harvard-Smithsonian Center for Astrophysics, 60 Garden Street, Cambridge, MA 02138, USA}

\author{W.~N.~Brandt}
\affiliation{\ignore{Eberly}Department of Astronomy and Astrophysics, Eberly College of Science, The Pennsylvania State University, 525 Davey Laboratory, University Park, PA 16802, USA}
\affiliation{Department of Physics, The Pennsylvania State University, 104 Davey Laboratory, University Park, PA 16802, USA}
\affiliation{\ignore{IGC}Institute for Gravitation and the Cosmos, The Pennsylvania State University, University Park, PA 16802, USA}


\author{D.~M.~Crenshaw}
\affiliation{\ignore{Georgia}Department of Physics and Astronomy, Georgia State University, 25 Park Place, Suite 605, Atlanta, GA 30303, USA}

\author{E.~Dalla~Bont\`{a}}
\affiliation{\ignore{Padova}Dipartimento di Fisica e Astronomia ``G. Galilei,'' Universit\`{a} di Padova, Vicolo dell'Osservatorio 3, I-35122 Padova, Italy}
\affiliation{\ignore{INAF}INAF-Osservatorio Astronomico di Padova, Vicolo dell'Osservatorio 5 I-35122, Padova, Italy}


\author{G.~De~Rosa}
\affiliation{Space Telescope Science Institute, 3700 San Martin Drive, Baltimore, MD 21218, USA}



\author{M.~M.~Fausnaugh}
\affiliation{Department of Astronomy, The Ohio State University, 140 W 18th Ave, Columbus, OH 43210, USA}
\affiliation{\ignore{}Kavli Institute for Space and Astrophysics Research,  Massachusetts Institute of Technology, 
77 Massachusetts Avenue, Cambridge, MA 02139-4307, USA}

\author{J.~M.~Gelbord}
\affiliation{Spectral Sciences Inc., 4 Fourth Ave., Burlington, MA 01803, USA}
\affiliation{Eureka Scientific Inc., 2452 Delmer St. Suite 100, Oakland, CA 94602, USA}

\author{M.~R.~Goad}
\affiliation{\ignore{Leicester}Department of Physics and Astronomy, University of Leicester,  University Road, Leicester, LE1 7RH, UK}


\author{A.~Gupta}
\affiliation{Department of Astronomy, The Ohio State University, 140 W 18th Ave, Columbus, OH 43210, USA}

\author{Keith~Horne}
\affiliation{\ignore{SUPA}SUPA Physics and Astronomy, University of St. Andrews, Fife, KY16 9SS Scotland, UK}

\author{J.~Kaastra}
\affiliation{\ignore{SRON}SRON Netherlands Institute for Space Research, Sorbonnelaan 2, 3584 CA Utrecht, The Netherlands}
\affiliation{\ignore{Leiden}Leiden Observatory, Leiden University, PO Box 9513, 2300 RA Leiden, The Netherlands}


\author{C.~Knigge}
\affiliation{\ignore{Southampton}School of Physics and Astronomy, University of Southampton, Highfield, Southampton, SO17 1BJ, UK}


\author{K.~T.~Korista}
\affiliation{\ignore{WM}Department of Physics, Western Michigan University, 1120 Everett Tower, Kalamazoo, MI 49008-5252, USA}

\author{I.~M.~M$^{\rm c}$Hardy}
\affiliation{\ignore{Southampton}School of Physics and Astronomy, University of Southampton, Highfield, Southampton, SO17 1BJ, UK}



\author{R.~W.~Pogge}
\affiliation{Department of Astronomy, The Ohio State University, 140 W 18th Ave, Columbus, OH 43210, USA}
\affiliation{Center for Cosmology and AstroParticle Physics, The Ohio State University, 191 West Woodruff Ave, Columbus, OH 43210, USA}

\author{D.~A.~Starkey}
\affiliation{\ignore{SUPA}SUPA Physics and Astronomy, University of St. Andrews, Fife, KY16 9SS Scotland, UK}
\affiliation{\ignore{Illinois}Department of Astronomy, University of Illinois Urbana-Champaign, 1002 W. Green Street, Urbana, IL 61801, USA}



\author{M.~Vestergaard}
\affiliation{\ignore{Dark}Dark Cosmology Centre, Niels Bohr Institute, University of Copenhagen, Vibenshuset, Lyngbyvej 2, DK-2100 Copenhagen \O, Denmark}
\affiliation{\ignore{Steward}Steward Observatory, University of Arizona, 933 North Cherry Avenue, Tucson, AZ 85721, USA}

\begin{abstract}
The Space Telescope and Optical Reverberation Mapping Project (AGN STORM) on NGC 5548 in 2014 is one of the most intensive multi-wavelength AGN monitoring campaigns ever. For most of the campaign, the emission-line variations followed changes in the continuum with a time lag, as expected. 
However, the lines varied independently of the observed UV-optical continuum during a 60-70 day “holiday,” suggesting that unobserved changes to the ionizing continuum were present. To understand this remarkable phenomenon and to obtain an independent assessment of the ionizing continuum variations, we study the intrinsic absorption lines present in NGC 5548. We identify a novel cycle that reproduces the absorption line variability and thus identify the physics that allows the holiday to occur. In this cycle, variations in this obscurer's line-of-sight covering factor modify the soft X-ray continuum, changing the ionization of helium. Ionizing radiation produced by recombining helium then affects the level of ionization of some ions seen by HST.  In particular, high-ionization species are affected by changes in the obscurer covering factor, 
which does not affect the optical or UV continuum, so appear as uncorrelated changes, a ``holiday''. It is likely that any other model which selectively changes the soft X-ray part of the continuum during the holiday can also explain the anomalous emission-line behavior observed. 
\end{abstract}


\section{INTRODUCTION}

In 2013, NGC 5548 was the subject of an intensive monitoring campaign based primarily on X-ray data from {\em XMM--Newton} and the 
{\em Neil Gehrels Swift Observatory}, supplemented with spectra from the {\em Hubble Space Telescope} Cosmic Origins Spectrograph
\citep{Kaastra14,Mehd15,Mehd16,Arav15,Ursini15,DiGesu15,Whewell15,Ebrero16,Cappi16}.
In the following year, an intensive ultraviolet and optical reverberation-mapping program (the Space Telescope and Optical Reverberation Mapping program, or AGN STORM) was undertaken using {\em Hubble Space Telescope} \citep{DeRosa15,Kriss18},
the {\em Neil Gehrels Swift Observatory} \citep{Edelson15}, ground-based telescopes for both imaging
\citep{Fausnaugh16} and spectroscopy \citep{Pei17}, and the 
{\em Chandra X-Ray Observatory} \citep{Mathur17}. This program yielded the first high-fidelity measurements of interband continuum lags in NGC 5548
\citep{Edelson15,Fausnaugh16,Starkey17}  and some very surprising emission-line results
\citep{Goad16,Pei17} --- in particular, some 60 days into the campaign, the broad emission lines stopped responding strongly to continuum variations. However, by the end of the six-month campaign, the normal relationship between the continuum and broad emission lines was restored. To the AGN STORM team, it appeared as the BLR had ``gone on a holiday" and for that reason, we will continue to refer to the period of anomalous emission-line response as the ``BLR holiday.'' 

It was subsequently noted \citep{Kriss18} that the behavior of the narrow absorption lines changed during the BLR holiday, with the lower-ionization lines continuing to track the observed UV continuum, but with only decorrelated changes in the \
higher-ionization absorption lines. 
These holidays are not a prediction of photoionization theory and the current standard model 
of the geometry of an AGN. Understanding the physics behind the holiday is essential because line-continuum reverberation is the only direct way to measure the mass of the central black hole, and this method is based on the existence of a correlation between the continuum and broad emission lines. 
The purpose of this paper is to begin an exploration of the BLR holiday phenomenon by examining the behavior of the narrow absorption lines, drawing extensively from the results obtained during the 2013 {\em XMM-Newton} program as well as the 2014 AGN STORM campaign. We focus on the absorption lines since the geometry is much simpler than the emission-line geometry. Absorbing gas must lie along our line of sight and the clouds must see the same SED as we do. Later sections discuss the observational constraints on the physics behind the holiday. We identify a novel physical process in which changes in the SED cause changes in the ionization of helium, which then drives the absorption line changes observed by {\em HST}. 

\section{THE GEOMETRY AND THE OBSCURER} 
\begin{figure}[H]
\centering
\includegraphics [width=4in]{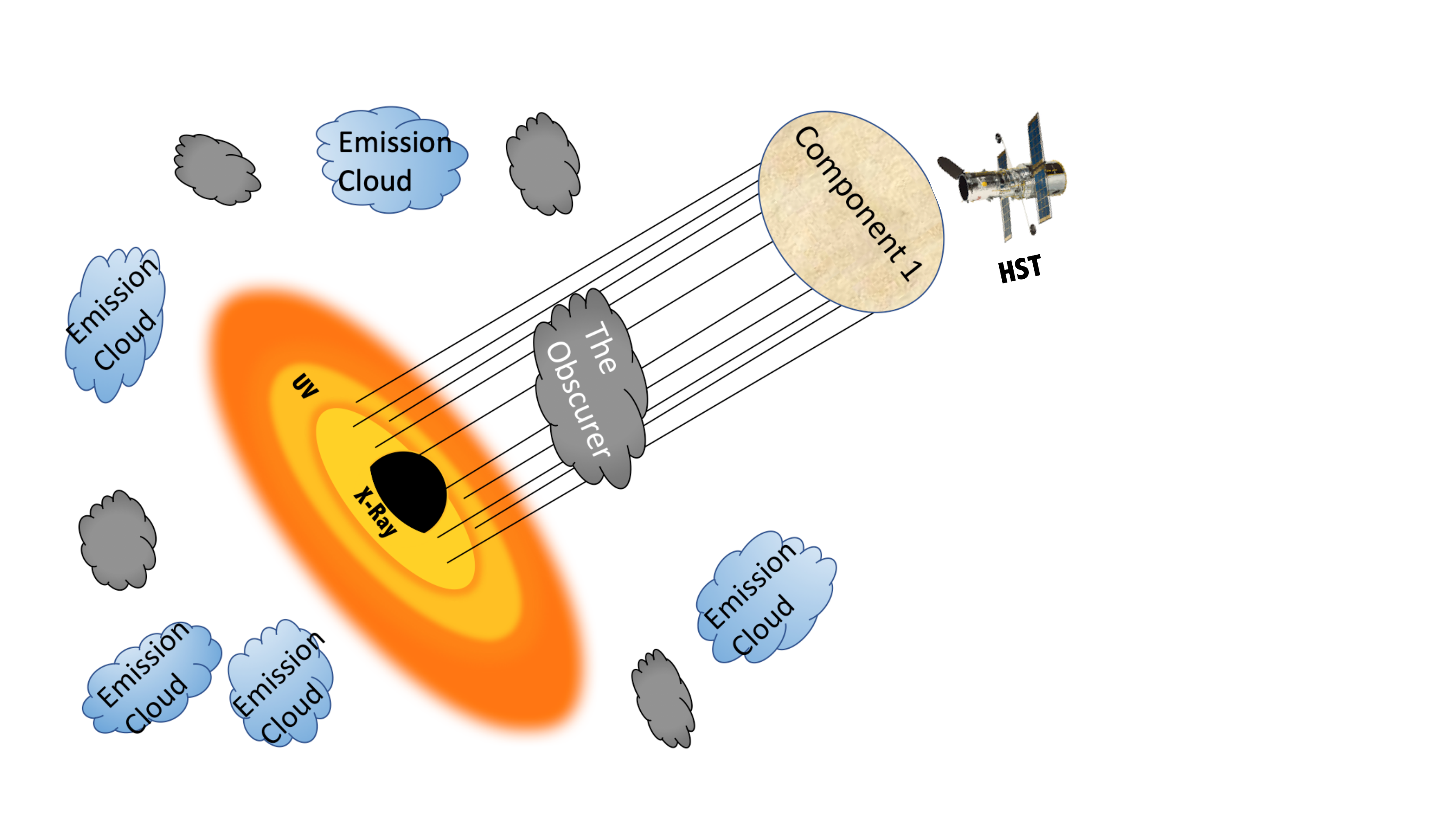}
 \caption{The geometry of the emission and absorption components discussed in this paper.  The line of sight obscurer covers 70\% to 100\% of the X-ray source \citep{Mehd16}. The gray clouds show other possible obscurers. These will be discussed in the next paper in this series. The blue blobs indicate the BLR clouds surrounding the source.  Component 1, the narrow absorption line component with the highest outflow velocity ($-1165~{\rm km~s}^{-1}$), is also shown. Component 1 is the absorbing component studied in this paper.   }\label{fig:f1}
\end{figure}

Historically, our line of sight to the central regions of NGC 5548 has been fairly clear, with no heavy obscuration in the soft X-ray, although ``warm absorbers'' are present \citep{Mathur95, Kaastra00}. Dramatic changes in the broadband soft X-ray absorption occurred and were interpreted as being due to a cloud, ``the obscurer,'' passing across our line of sight
\citep{Kaastra14,Mehd16}.  This kind of heavy obscuration had never been seen before in NGC 5548 although a similar extinction occurred in NGC 4151 \citep{Ferland82}. Soft X-ray absorption by the obscurer was first observed in NGC 5548 in 2012 and 2013
\citep{Mehd16,Arav15}.  Here we briefly summarize the geometry inferred by the {\em XMM-Newton} ``Anatomy'' series of papers \citep{Kaastra14,Mehd15,Mehd16,Arav15,Ursini15}. 
Figure~\ref{fig:f1} shows a sketch of the overall geometry, including the black hole and accretion disk which produce the intrinsic (unobscured) SED. The observer is located in the direction of the {\em HST} icon.
  

A set of six intrinsic, narrow
FUV\footnote{We refer to the region 6 -- 13.6 eV (912\,\AA\ to 2000\,\AA)
as FUV; 13.6 -- 54.4 eV (228\,\AA\ to 912\,\AA) as EUV; and 54.4 eV
to few hundred eV (less than 228\,\AA) as XUV.
absorption lines potentially associated with
the X-ray warm absorber in NGC 5548 \citep{Mathur95} have been studied in detail
by \cite{Mathur99}, \cite{Cren03}, \cite{Arav15} and \cite{Kriss18}.
They are numbered from 1 to 6 in order of decreasing outflow
velocity from $-1165~{\rm km~s}^{-1}$ to $+250~{\rm km~s}^{-1}$.
In this paper, we specifically study the cloud producing Component 1, which is
also illustrated in Figure~\ref{fig:f1}.
Component 1 is the closest narrow FUV absorber to the central source,
and it is the target of our study since it shows the most dramatic changes
during the obscuration \citep{Arav15}.
This component also has the greatest assortment of associated UV absorption
lines, permitting its characteristics to be determined in detail.
The full physical properties of this component were derived by \cite{Arav15}.
In particular, it has a density of log $n_e = 4.8 \pm 0.1~{\rm cm}^{-3}$,
as measured using the metastable absorption lines of \ion{C}{3} and \ion{Si}{3}.
This leads to a well determined distance of $3.5 \pm 0.1$ pc,
placing it outside the BLR but within the narrow-line region (NLR)
\citep{Pet13}.}

Little is known about the density and location of the obscurer,
but the ionization state, inferred high density, the kinematics,
absorption line profiles, and the covering factors all suggest an origin
in the BLR \citep{Kaastra14,DiGesu15,Mehd15}.
BLR lags of two days to $\sim$ ten days would imply a distance of
between $6 \times 10^{15} - 3 \times 10^{16}$\,cm \citep{DeRosa15}.
The soft X-ray observations show that the obscurer does not fully cover the
X-ray source.
The covering factor varied between 0.7 and 1.0 over the period
2012 to 2015 \citep{Mehd16}.
This change may be caused by either transverse motions of the obscurer,
changes in its internal structure, or
changes in the size of the X-ray source \citep{Mehd16}.
Finally, it is possible that other obscurers lie within the central regions,
as shown in Figure~\ref{fig:f1}.

\section{THE HOLIDAY}
In photoionization equilibrium, there is a correlation between the brightness of the ionizing radiation field and the ionization state of the gas. Reverberation measurements rely on this, with the only complication being the time lag caused by the finite speed of light. ``Holidays,''  where the correlation breaks down, are not expected. This section outlines the absorption and emission line holidays that occurred during the AGN STORM campaign. 

\subsection{Broad emission lines and their holiday} 
The AGN STORM campaign monitored NGC 5548 for the 6-month period from 2014 January to July. During almost 120 days of the campaign, the BLR emission exhibited the expected correlation with the continuum. The line and continuum emission holiday started about 75 days after the first HST observation and continued for 60 to 70 days \citep{Goad16}. The lines then returned to their normal behavior. \cite{Goad16} and \cite{Pei17} note that the strong and broad emission lines became significantly fainter (e.g. in CIV and H$\beta$) during the holiday. This is the best-characterized anomalous event detected in a reverberation mapping campaign. 

\subsection{Narrow absorption lines and their holiday} 
\begin{figure*}
\centering
\includegraphics [width=6in]{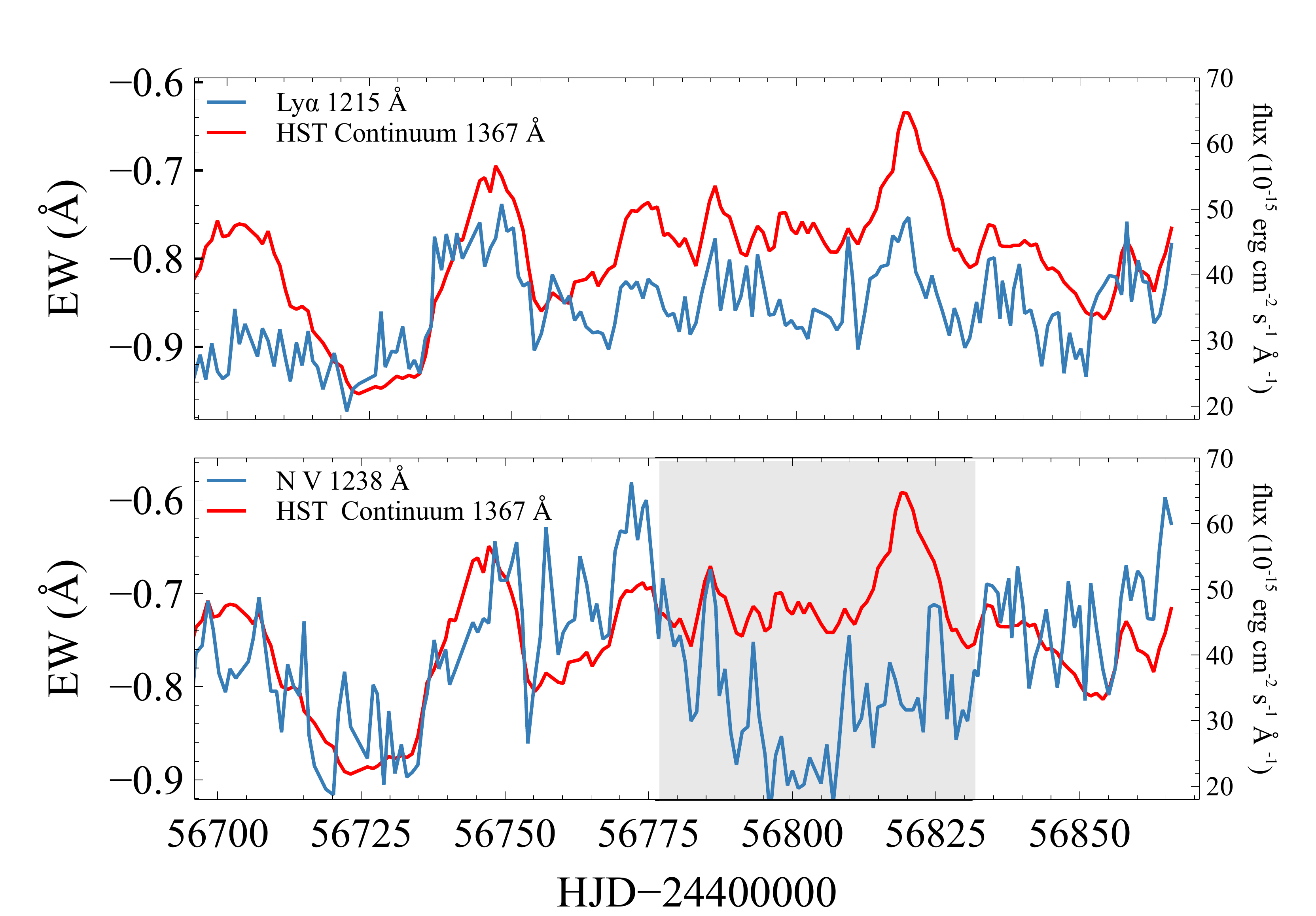}

\caption{Both panels show the arbitrarily scaled FUV continuum in red, as a function of Heliocentric Julian Date $-$24400000. 
The upper panel shows the equivalent width of the \lya\
absorber of Component 1 in blue and the lower panel shows the equivalent with of the corresponding \nv\,$\lambda1238$ absorption line. Shaded area indicates the time in which the "holiday" is happening.}\label{f2}\end{figure*}

Some, but not all, of the Component 1 absorption lines displayed a holiday similar to the emission lines. Three low-ionization species  ---
H\,{\sc i}, \siII, and \cii\ --- showed good correlations with the {\em HST} FUV continuum, while the higher-ionization species ---
\siIII, \siIV, \ciii, \civ, and \nv\  --- showed decorrelated behavior \citep{Kriss18}. Figure~\ref{f2} shows examples of both behaviors, \lya\ and 
\nv\,$\lambda1238$. The red line shows the arbitrarily  scaled {\em HST} FUV continuum while the blue lines are Component 1 absorption line equivalent widths (EW). Both lines correlate for most of the campaign, but, like the broad emission lines, there is an almost 70-day period when \nv\ is decorrelated. 

\subsection{The scope of this paper}

This paper focuses on the absorption line holiday with the aim of reproducing 
the correlation / decorrelation shown in Figure~\ref{f2}.
As Figure\ref{fig:f1} shows, the absorption lines involve an especially simple geometry with the continuum emitters and absorbers lying along a single line of sight and the absorbing clouds being illuminated by the SED directed towards the Earth. 
The absorption lines vary due to changes in the SED striking Component 1. These changes may be due to 
variations of the brightness of the AGN or changes in the shape of the SED 
caused by changes in the obscurer's absorption.  

The emission lines are more complicated. They might not be directly affected by the obscurer since they lie along different sight lines from the AGN 
(Figure~\ref{fig:f1}).  Other obscurers may be present on other sight lines and could affect the emission lines. 
The emitting clouds have a range of densities and distances from the center. 
All of this introduces complexities. 
The absorption lines are the simplest, and so the best place to start studying the physics behind the holiday. 
They are the focus of this paper. 

The remainder of the paper examines how an absorption line holiday can occur. 
Our goal is to identify a physical process whereby lines sometime
correlate with the observed FUV continuum, and at other times do not.
We aim to identify the phenomenology that makes this possible, but 
not model any particular HST observation.  Converting between observed equivalent widths
and the ionic column densities we predict requires a curve of growth
analysis.
This brings in additional uncertainties including the velocity field of
the gas and possible substructure within the absorption lines.  
The presence or absence of a correlation between the FUV continuum and 
a line equivalent width or column density
will not be affected by curve of growth effects.
In other words, we want to reproduce the correlation / decorrelation of the lines
and continuum and do not model specific derived column densities. 
We build upon the \citet{Arav15} model of Component 1 and do not change its basic assumptions. In the following section we present the standard parameters for Component 1, taken from \citet{Arav15}.

\section{THE ``STANDARD'' MODEL OF COMPONENT 1}

\begin{figure*}
\centering
\includegraphics [width=6 in]{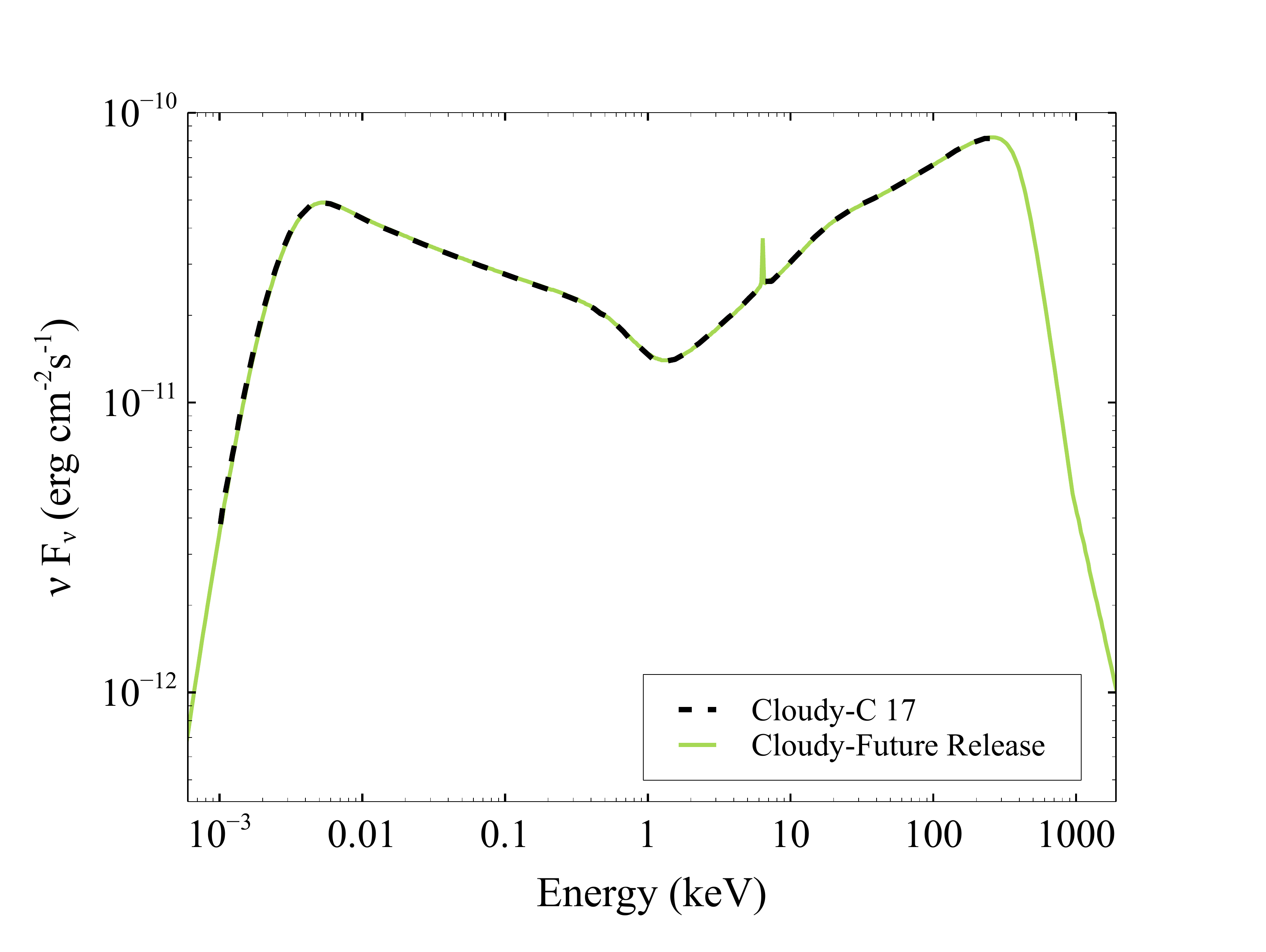}
\caption{The intrinsic (unobscured) SED available in version 17 of {\tt cloudy} (C17) is shown in dashed-style black line. It was zero outside the indicated range. The green line shows the improved SED (\cite{Mehd15}) which will be implemented in future versions of {\tt cloudy}.}\label{f3}
\end{figure*}

Below we use photoionization models to investigate why some absorption lines correlate with the FUV continuum and some do not. We first adopt the intrinsic SED emitted by the accretion disk, shown in Figure~\ref{f3}. This was derived by continuum modeling during the multi-wavelength campaign data on NGC 5548 \citep{Mehd15} and is used in all calculations presented below. 
This is based on a Comptonized thermal accretion disk model.   This model was derived using simultaneous, coordinated multiwavelength observations including FUV and soft X-ray \citep{Kaastra14} and hard X-ray continua \citep{Ursini15}.
This SED was incorporated into the developmental version of {\tt cloudy}, most recently described by 
\cite{Ferland17}, and will be available in the next release. We use this developmental version throughout this paper. Version 17, the latest public release of {\tt cloudy}, included an NGC 5548 SED derived by Tek P.\ Adhikari from CAMK (Warsaw), by digitizing figure 10 of 
\cite{Mehd15}. That SED did not include data for energies not included in the published figure. The improved SED used by 
\cite{Mehd15} covers the entire electromagnetic spectrum, and includes the observed Fe K$\alpha$ line. 

The principal conclusion of \cite{Hop04} is that extinction in SDSS quasars is typically E(B-V) =0.013 mag with an SMC-like extinction curve. There are, of course, “red quasars” that are heavily extinguished \citep{Gas17},but NGC 5548 is not one of them. There are hard X-ray observations with NuSTAR and INTEGRAL observatories, and the SED model that we use matches the observed flux and shape of the hard X-ray continuum \citep{Ursini15}. The SED model that was derived by \cite{Mehd15} is fully consistent with simultaneous observations taken with multiple observatories from optical to hard X-rays.

\begin{figure*}
\centering
\includegraphics [width=\textwidth]{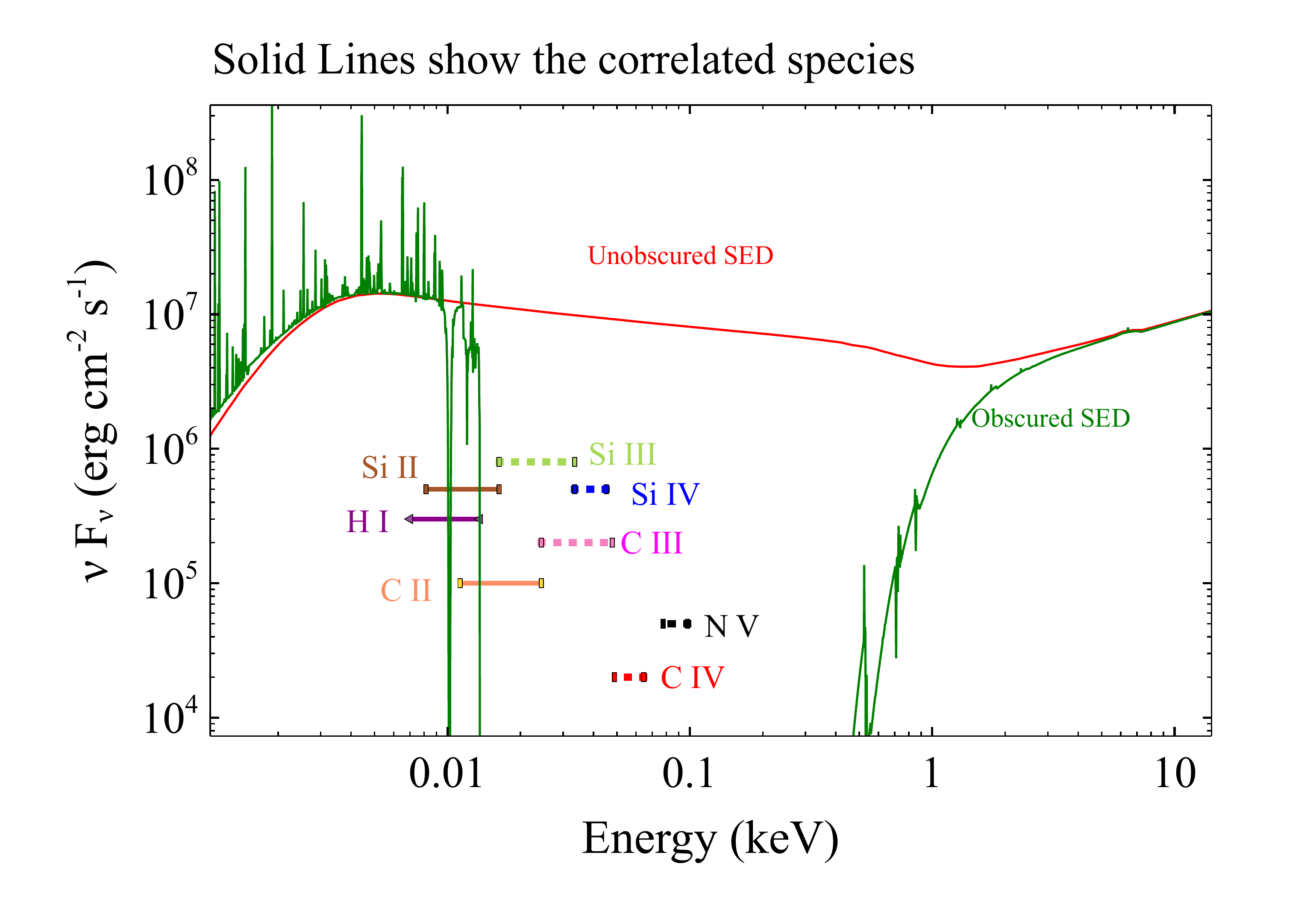}
\caption{The expected SED  transmitted through the obscurer and striking Component 1. Each line segment shows the energy required to produce the ion (left end) and to destroy the ion (right end). In this Figure, we assumed that the obscurer fully covers the continuum source. The column density to make the transmitted  SED is $N({\rm H})=1.2\times10^{22}\,{\rm cm}^{-2}$}. \label{f4}
\end{figure*}

We adopt the obscurer parameters --- $N({\rm H})=1.2\times10^{22}\,{\rm cm}^{-2}$ and $\log \xi = -1.2$ (erg cm s$^{-1}$) --- derived by 
\cite{Kaastra14}. The ionization parameter $\xi$ is defined as \citep{Tarter69, Kallman01}
\newline
\begin{equation}
\xi= \frac{L}{n({\rm H}) R^2},
\end{equation}
where $L$ is the luminosity of the ionizing source over the 1--1000 Ryd (13.6 eV to 13.6 keV) band in erg s$^{-1}$, $R$ is the distance from the source in cm. For the hydrogen density, we adopt $n({\rm H})= 10^{10}$ cm$^{-3}$ , which is a typical BLR cloud density. We adopt the {\tt cloudy} default value\footnote{More information about the default values can be find in HAZY: \href{url}{https://www.nublado.org/wiki/DownloadLinks}} for solar abundances, which are generally within 30\% of the \cite{Lodders03} meteoritic abundances used in some of the previous modeling. 
The transmitted SED calculated with these parameters is shown in Figure~\ref{f4}, in which we assume that the obscurer fully covers the continuum source. This figure shows the net
transmitted radiation field at the shielded face of the obscurer.  It includes the attenuated
incident radiation field produced by the central object along with line and continuum emission
produced by the obscurer. The effects of filtering the continuum has been discussed in other literature but within different contexts \textcolor{red}{\textbf{\citep{Ferland82,le04}}}. 
The horizontal lines in Figure~\ref{f4} indicate the ionization energies for the species studied by \cite{Kriss18}. The left terminus of each line shows the energy needed to produce the ion, while the right terminus indicates the amount of energy required to destroy the ion by further ionization. The high ionization-potential species (indicated by dotted lines) did not correlate with the FUV during the holiday, while lower ionization potential species (solid lines) remained correlated.

In the case of Component 1, we adopt the parameters from \cite{Arav15}, $\log n({\rm H}) = 4.72$ (cm$^{-3}$) and ionization parameter
$\log U = -1.5$, which is defined to be \citep{Osterbrock06}: 
\begin{equation}
U = \frac{Q({\rm H})}{4 \pi R^2 n({\rm H}) c},
\end{equation}
where $Q({\rm H})$ is the number of hydrogen-ionizing photons emitted by the source per second, $R$ is the cloud distance from the ionizing continuum source, which is 3-5 pc for Component 1 as described in section 2. In the above equation $c$ is the speed of light. 
Note that the papers modeling the obscurer \citep{Kaastra14} and Component 1 \citep{Arav15} use different definitions for the ionization parameter. For the unobscured SED, the relation $\log U = \log \xi -1.6$ can be used to convert between these ionization parameters. For the obscured SED (Figure~\ref{f4}, green line), the conversion relation is $\log U= \log \xi - 3.3$.

\section{ WHAT HAPPENED?} 
We hypothesize that two independent events occurred. First, the luminosity of the AGN varied, causing the entire SED to become brighter or fainter. This would cause the expected correlated variations. Second, the obscurer moved across our line of sight, perhaps due to its orbital motion around the black hole, changing the fraction of the central source that is covered. We will show below that absorption by the obscurer changes the EUV, XUV, and soft X-ray portions of the SED but has little effect on the optical, UV, or FUV, where the obscurer is transparent. This variable absorption, caused by the changing covering factor, would affect the high ionization absorption lines but have little effect on the FUV or optical continuum, so would produce decorrelated changes i.e., a "holiday". In the rest of this section, we investigate these two events in more detail. 

\subsection{Changing the luminosity of the source} 
\begin{figure*}
\centering
\includegraphics [width=6in]{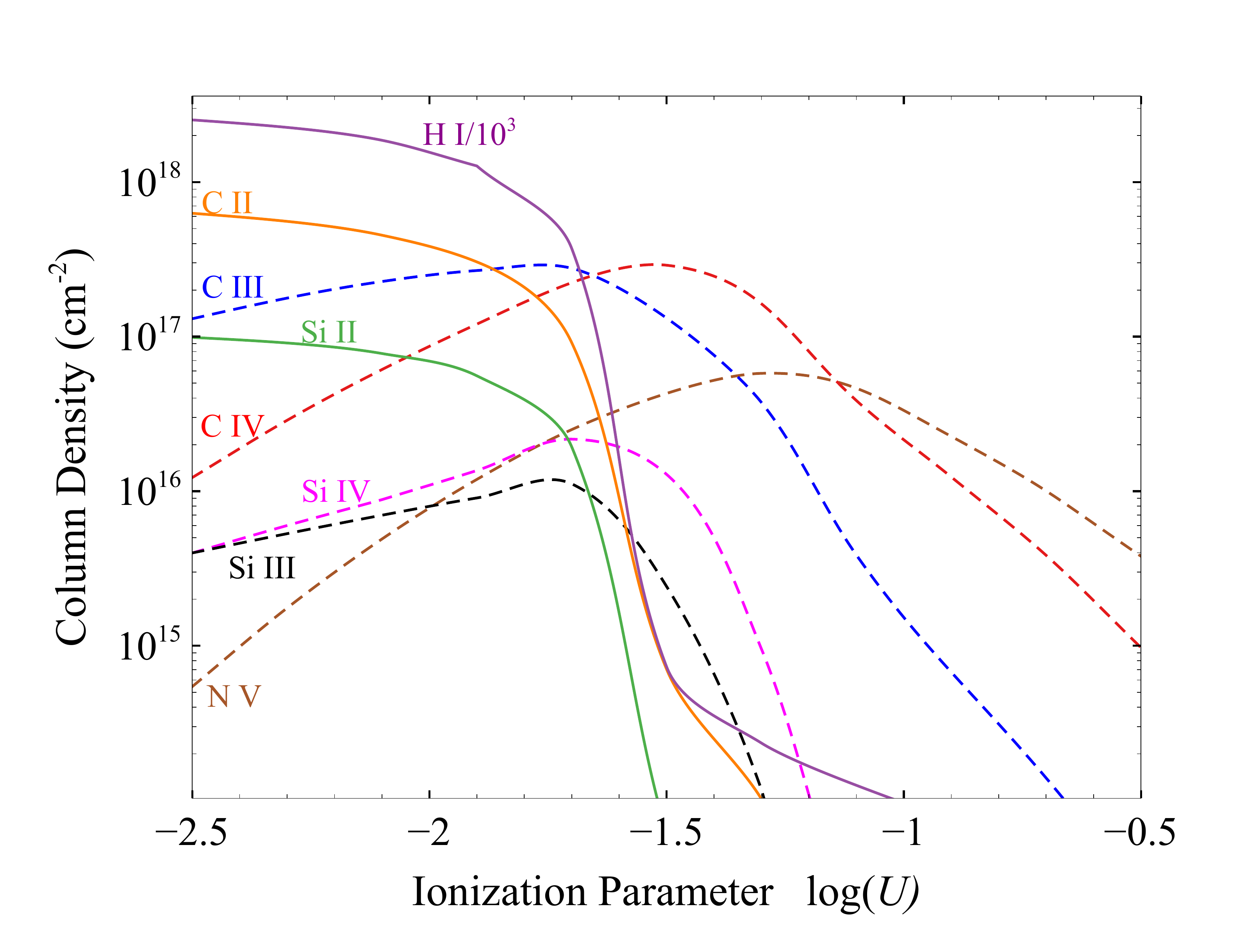}
\caption{This shows how Component 1 column densities change as the ionization parameter changes. The 
hydrogen column density is divided by 1000 for comparison purposes. The solid lines are correlated species while the 
dashed lines are decorrelated. These changes are unlike those seen in the holiday, ruling out changes in $U$ alone as the 
reason for the holiday.}\label{f5}
\end{figure*}

The changing luminosity is directly seen via optical, FUV, and X-ray observations. In photoionization equilibrium, this implies a varying ionization parameter, which would change the column densities of all species. 
As a test, we checked what happens to the column densities of Component 1 absorbing species when the continuum luminosity changes, while keeping the unobscured shape of the SED the same. We use the 
\cite{Arav15} standard Component 1 parameters as described above, but we let the ionization parameter $U$ vary by one dex to either side of the standard value of $\log U = -1.5$. This would correspond to changes in the continuum luminosity by the same amount. For comparison, the 1157\AA \hspace{2mm}{\em HST} continuum varied
    over a range of 0.6 dex during the STORM campaign. Figure~\ref{f5} shows the results of these calculations. The solid lines show the correlated species while the dashed lines are decorrelated. All the column densities change dramatically, however around the standard value of $\log U = -1.5$, the columns of the correlated species are changing very fast, and faster than those of the decorrelated ones. This is not enough to explain the holiday. Thus, simple changes in the luminosity of NGC 5548 cannot explain the absorption line holiday. We must look elsewhere. 

\subsection{Changing the obscurer covering factor}
\begin{figure*}
\centering
\includegraphics [width=6in]{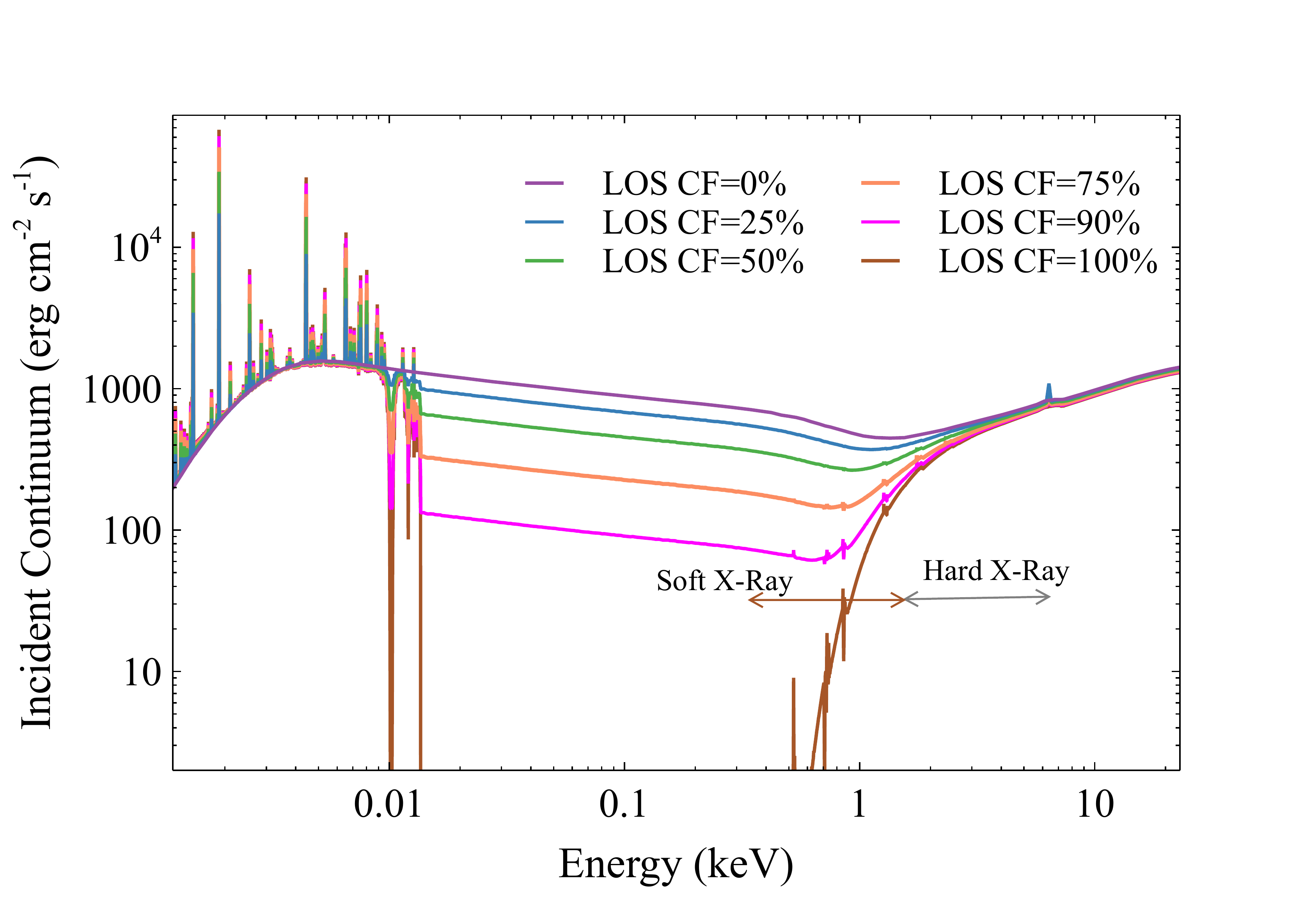}
\caption{The variations of the SED striking Component 1for different LOS CFs. The \cite {Arav15} ionization parameter is reproduced at zero coverage. The Figure indicates the soft and hard X-ray energies, i.e., 0.3--1.5 keV and 1.5--10 keV \citep{Mehd16}.}\label{f6}
\end{figure*}

The soft X-ray extinction measures the fraction of the continuum source covered by the obscurer. We refer to this as the ``line of sight covering factor'' (LOS CF). Changes in the LOS CF affect the absorption lines seen with {\em HST} since the SED transmitted through the obscurer is responsible for the ionization of Component 1. 
Figure~\ref{f6} shows how changes in the LOS CF affects ${\rm SED}_{\rm inc}$, the SED striking Component 1.  This is defined as:
\begin{eqnarray}
\nonumber
 {\rm SED}_{\rm inc} & = & ({\rm LOS\ CF)} \times  {\rm (SED_{extinguished})}\\
 & &  + (1 - {\rm LOS\ CF}) \times {\rm SED}
 \end{eqnarray}
for various LOS CF. Here “SED” indicates the unattenuated SED shown in Figure~\ref{f3}. The intensity is adjusted to $\log U =-1.5$ with LOS CF = 0
\citep{Arav15}.  We keep the brightness of ${\rm SED}_{\rm inc}$ constant at 4558\,\AA\ (0.2 Ryd), and vary the LOS CF to obtain different shapes. We chose the energy 
0.2 Rydberg since this is an energy where the obscurer is transparent. In Equation (3), ${\rm LOS\ CF} = 0$ will be the full unattenuated SED and 100\% coverage would be the \cite{Mehd15} 
extinguished SED (Figure~\ref{f4}). Figure~\ref{f6} shows that the 1 keV X-ray absorption is highly 
affected by changes in the LOS CF. The hard X-rays are not absorbed and so do not change. Note that this assumes that the LOS CF is the same for the EUV and XUV, whereas these components may form in different regions \citep{Gardner17,Edelson18}. 
Note that the SEDs shown in Figure~\ref{f6} are the \textit{incident} radiation field striking
the illuminated face of Component 1. The data come from the second column of the \texttt{cloudy} \texttt{save continuum}. 
The effects of diffuse fields from the obscurer are included when generating the extinguished SED.

\begin{figure*}
\centering
\includegraphics [width=6in]{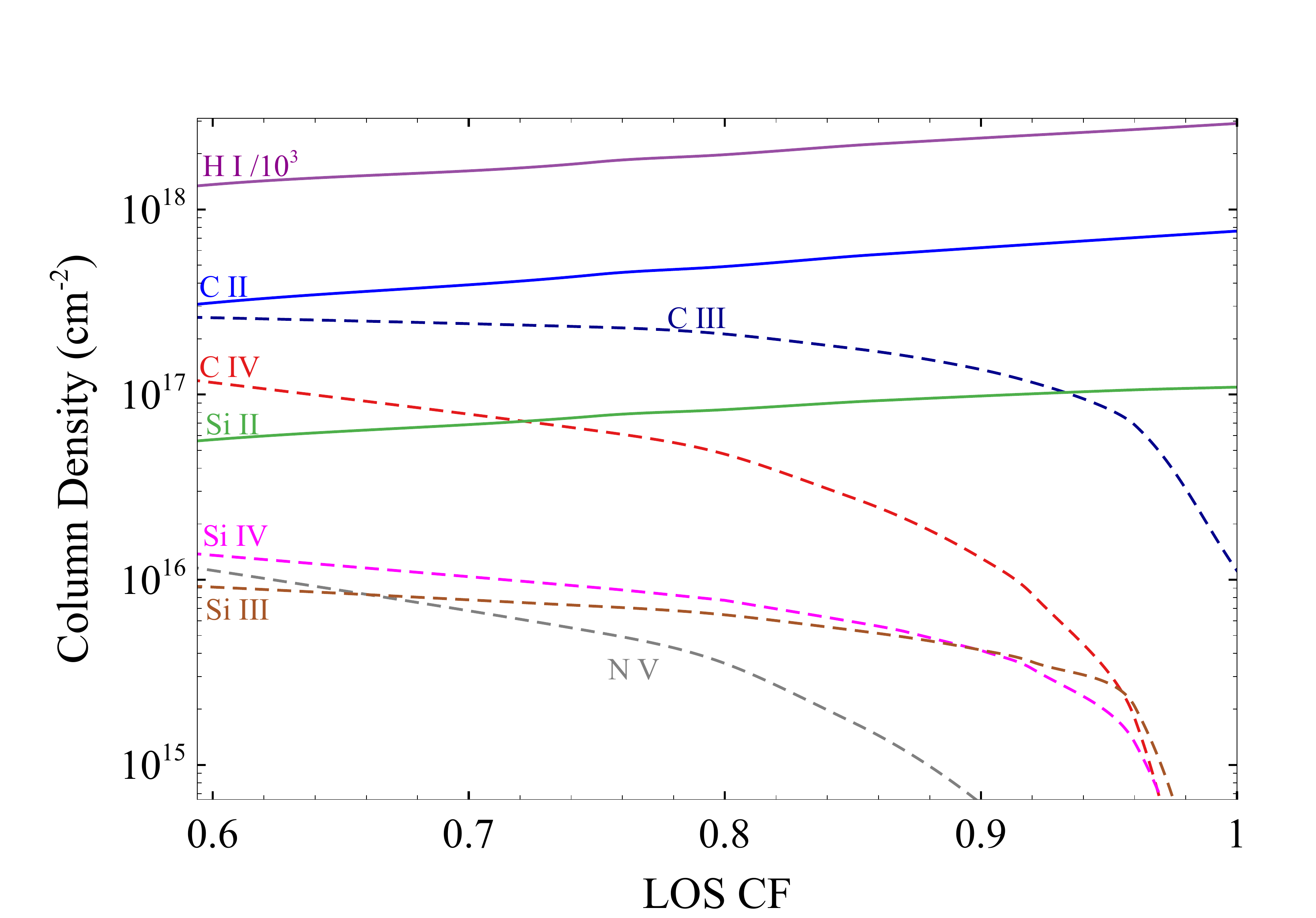}
\caption{The effects of changes in the obscurer LOS CF upon the column densities observed by {\em HST}. Low-ionization species (solid lines) remained correlated while high ionization (dashed) species were decorrelated during the holiday, as expected from changes in the obscurer covering factor. The ionization parameters adopted for the obscurer and Component 1 are $\log \xi = -1.2$ (erg cm s$^{-1}$) and $\log U = -1.5$, respectively.}\label{f7}
\end{figure*}

Variations of the obscurer LOS CF produce considerable changes in the transmitted SED without producing observable changes in the FUV since the obscurer is transparent in the FUV (see Figure~\ref{f6}). Perhaps this can provide an explanation for the correlated and decorrelated behavior of the narrow absorption lines of Component 1. Next, we investigate how the column densities of Component 1 are affected by the changes in the obscurer LOS CF. 
We used SEDs like those illustrated in Figure~\ref{f6} to predict the column densities of the Component 1 species measured by 
\cite{Kriss18}. These are shown in Figure~\ref{f7}. 
The column densities of high ionization species decrease while low-ionization species change little at high LOS 
CF values. Clearly, then, changes in the obscurer LOS CF are capable of causing the absorption line holiday. The 
next section outlines the physics behind Figure~\ref{f7}. 

\section{PHYSICS BEHIND THE ``HOLIDAY''}

\begin{figure*}
\centering
\includegraphics [width=7in]{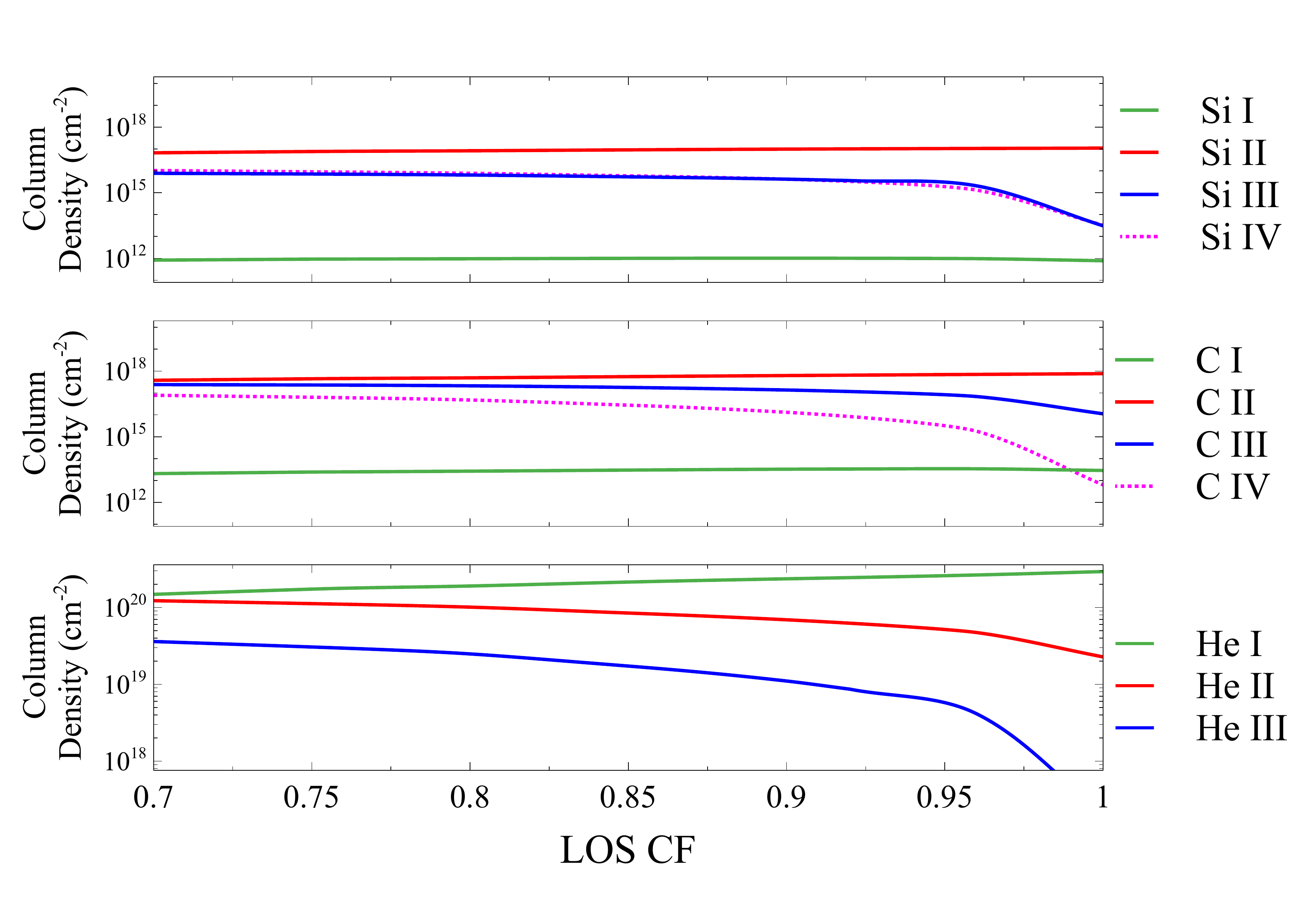}
\caption{Variations of the column densities of different ionization stages as the obscurer LOS CF changes. Si, C, and He are shown in three panels, from top to bottom, respectively. }\label{f8}
\end{figure*}

Figure~\ref{f7} focused on the absorption line species observed in the {\em HST} spectra. These are not necessarily the dominant or most important ions. Figure~\ref{f8} shows how the physically important ions change, and includes helium, which {\em HST} did not observe. 
Silicon and carbon are mainly singly ionized, while He is mostly neutral.

\begin{figure*}
\centering
\includegraphics [width=7in]{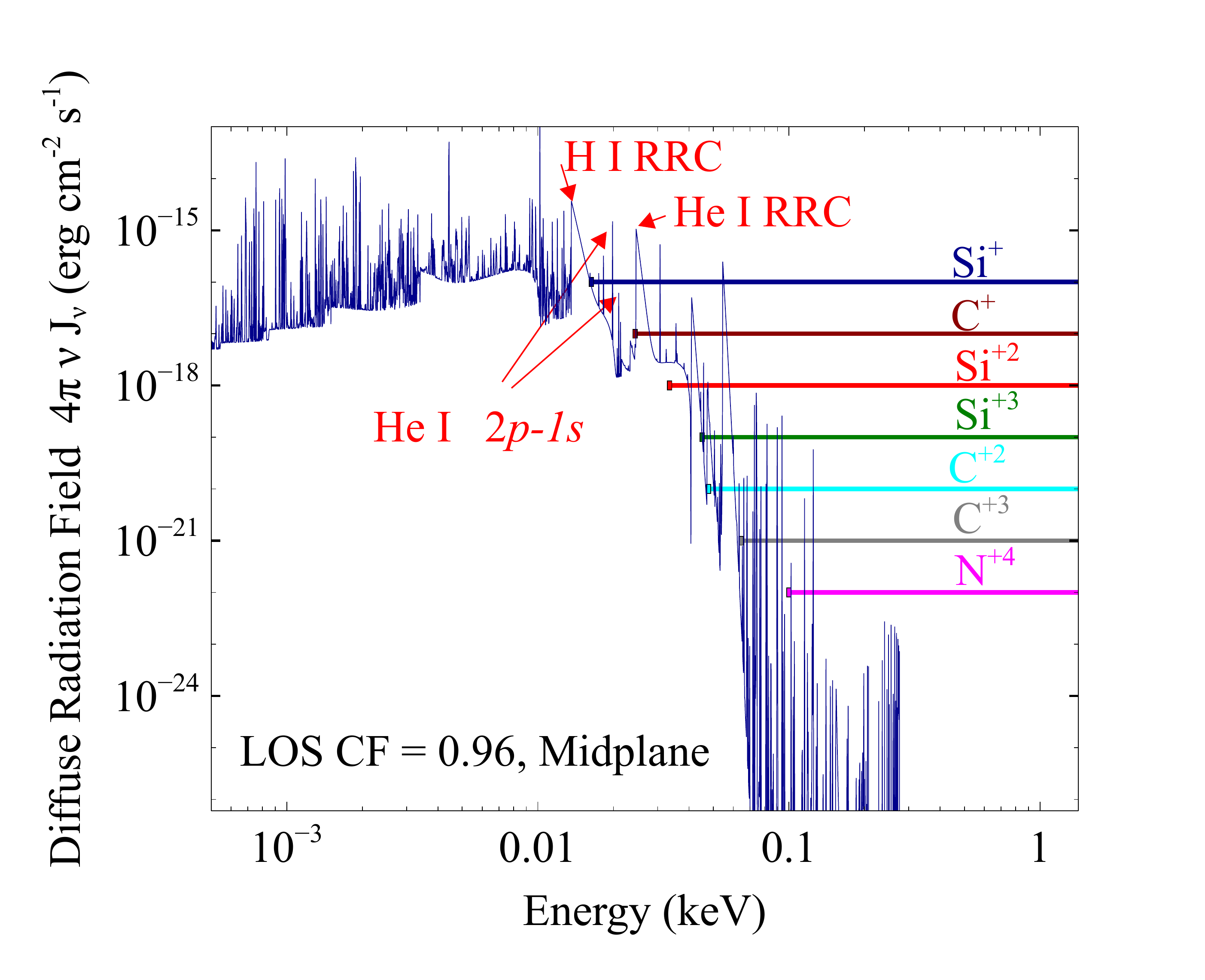}
\caption{Diffuse emission field at the midplane of Component 1 with 96\% obscuration. The left terminus of the horizontal lines shows the minimum energy needed to destroy the ion and produce the higher stage ion. RRC stands for radiative recombination continua and H I RRC is the Lyman continuum emission} .\label{f9}
\end{figure*}

As the LOS CF increases, the column densities of the higher ionization-potential decorrelated ions decrease dramatically, as also seen in Figure~\ref{f7}. The ions He$^+$ and He$^{+2}$ behave like the higher ionization-potential decorrelated species. Si$^{+}$ and C$^+$ are the most abundant ions, and their column densities change only slightly.  
To understand this behavior, we must isolate what photoionizes the dominant and correlated low-ionization species to produce the decorrelated behavior in the higher ionization species. To answer this, we consider the radiation field within Component 1. Figure~\ref{f9} shows the diffuse radiation field at the midplane, the middle of the Component 1 cloud. The midplane is a representative location, and its properties give insight into the physics of the cloud. We chose a LOS CF of 96\%, which is representative of the regions of Figure~\ref{f7} where the correlated/decorrelated behavior is pronounced. This covering factor is so large that the EUV and XUV portion of the incident SED shown in Figure~\ref{f6} is faint.  The diffuse radiation field shown is produced by emission from the absorbing gas itself and is dominated by line and continuum emission produced by recombining helium. Several of the 
prominent emission features are labeled. The horizontal lines indicate the range of photon energy that 
can photoionize the indicated species.  

\begin{figure*}
\centering
\includegraphics [width=7in]{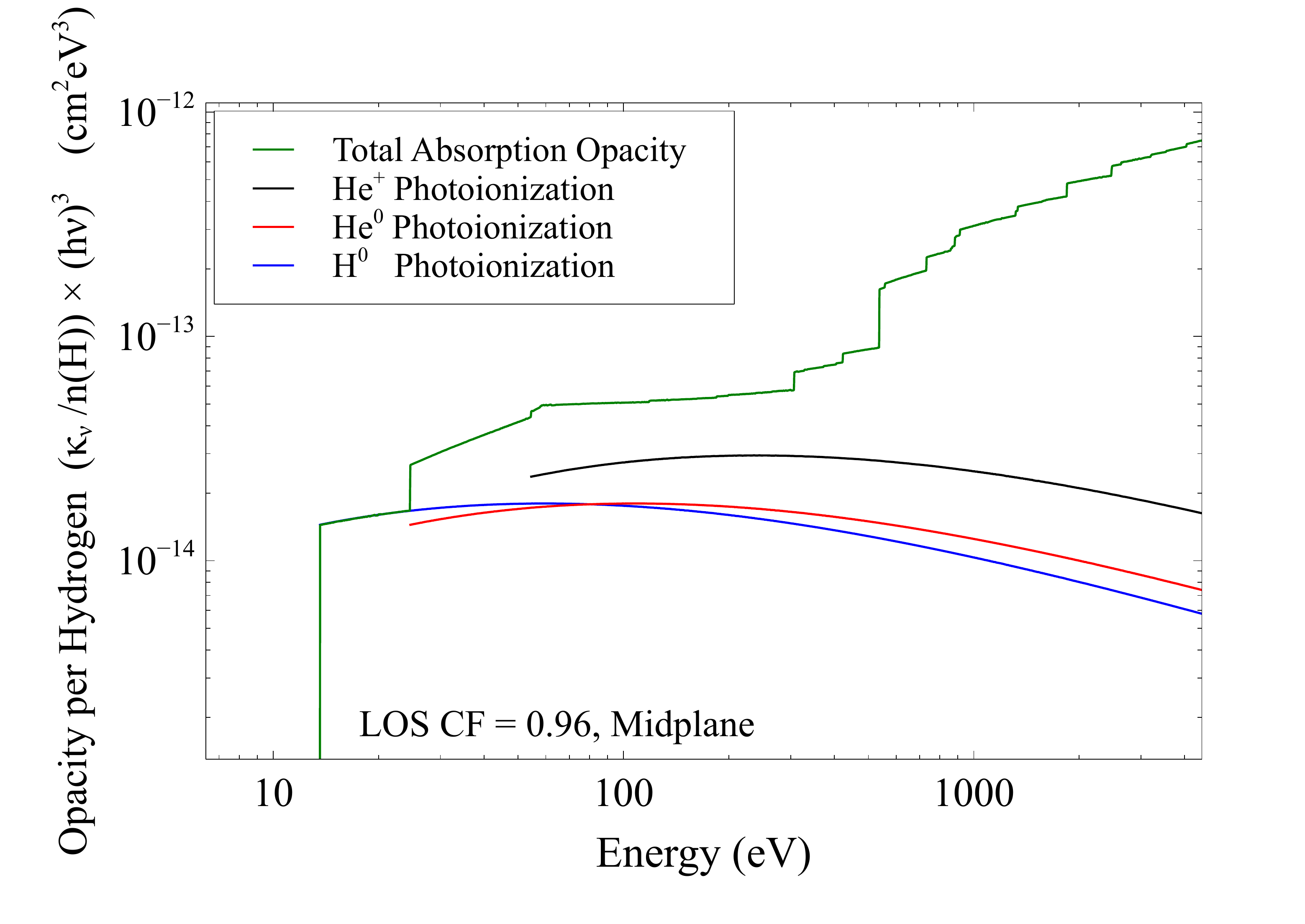}
\caption{Opacity per hydrogen atom is shown as a function of energy. This shows the total opacity at the Component 1 midplane. The vertical axis has been scaled by $h\nu^{3}$ for clarity, as described in the text.}\label{f10}
\end{figure*}

Examination of the photoionization rates shows that C$^+$ and Si$^+$ are produced by photoionization of the neutral atoms by the Balmer continuum. They are destroyed by valence-shell photoionization, with thresholds of 24.4 eV and 16.3 eV for C$^+$ and Si$^+$, respectively. Inner shell photoionization by the soft X-rays is much less important. The He\,{\sc i} radiative recombination continua (RRC) (for C$^+$ and Si$^+$) and singlet and triplet $2p-1s$ transitions of He$^0$ (for Si$^+$) are the primary sources of photoionization at energies of $\sim20$ -- 25 eV, the threshold for destroying these dominant species. These are all produced by recombination of He$^+$. This means that the abundances of the decorrelated 
high-ionization species follow the abundance of He$^+$ and subsequent He\,{\sc i} emission. Figure~\ref{f8} shows that the decrease in column density of the decorrelated species tracks changes in the He$^+$ column density. 
What is responsible for photoionization of He$^0$, producing He$^+$? He$^0$ is the dominant ion stage in Component 1 (Figure~\ref{f8}). Examination of the contributors to the photoionization rates shows 
that He$^+$ is produced through photoionization by soft X-rays from the attenuated SED of the AGN. 
Figure~\ref{f9} shows only the diffuse fields and does not include the attenuated incident SED. 
He$^0$ is an important opacity source for soft X-rays. Figure~\ref{f10} shows the continuous opacity at the midplane of the Component 1 cloud. We evaluated the total gas opacity for the predicted distribution of ions and the assumed solar composition. This shows the opacity per hydrogen and is multiplied by the cube of the photon energy so that it can be compared with standard plots of the total ISM opacity \citep{Ride77}. The green line shows the total opacity while the other lines show some of the important contributors to it. H$^0$ is dominant in the low-energy EUV, He$^0$ is dominant in the high-energy EUV and XUV, and the heavy elements dominate around 0.5 -- 1 keV 
(\citealt[][their figure 2]{Cruddace74,Ride77}), causing the stepped rise in the right part of the diagram.
Helium is mainly neutral (Figure~\ref{f8}) and Figure~\ref{f10} shows that helium is a major contributor to the total opacity for energies from 24 to  300~eV. 

\begin{figure*}
\centering
\includegraphics [width=5in]{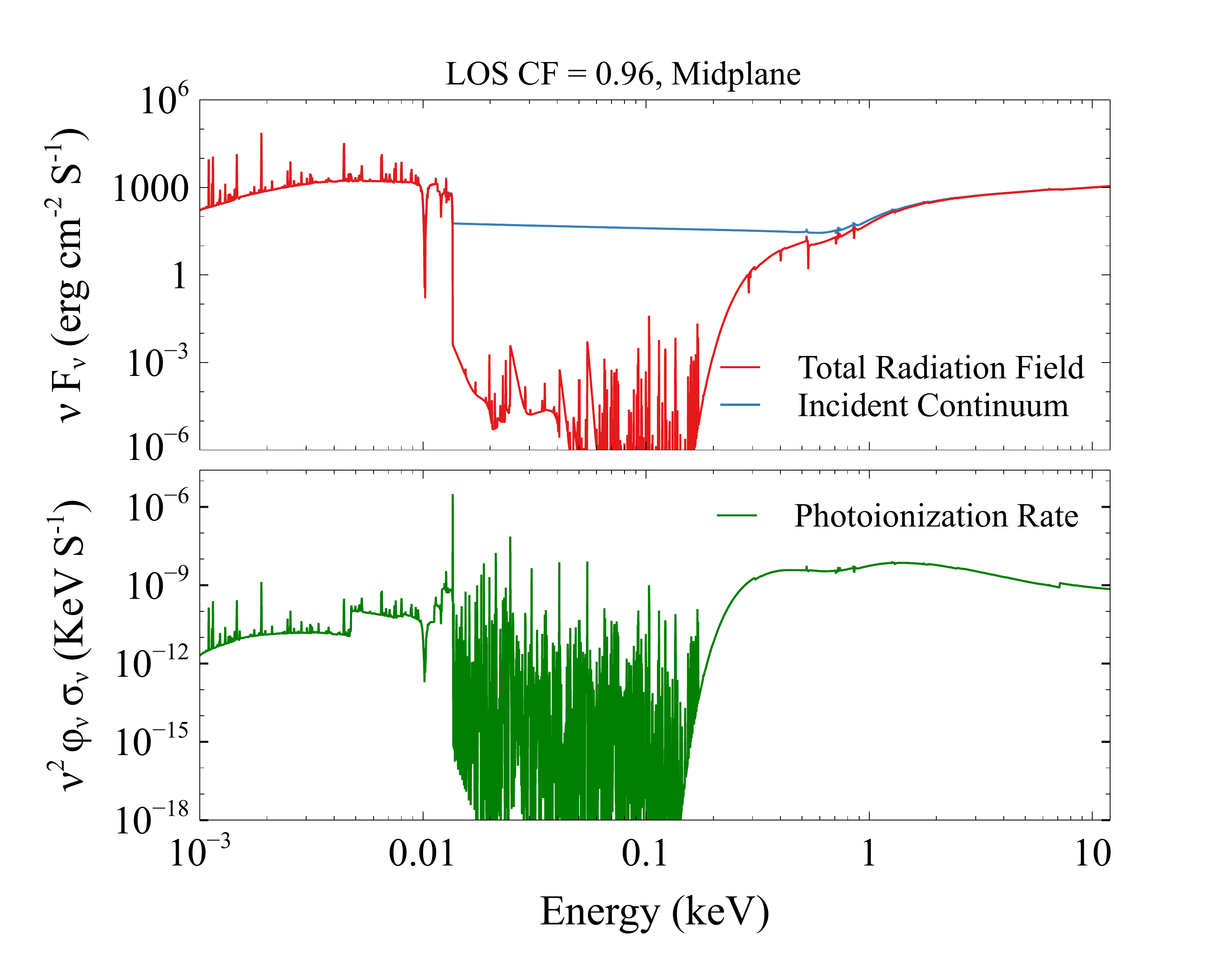}
\caption{The upper panel shows the radiation field striking Component 1 as the blue line and the radiation field at the midplane
of the cloud as the red line. The lower panel shows the photoionization rate at each energy.}\label{f11}
\end{figure*}

The gas photoionization rate is the integral of the opacity shown in Figure~\ref{f10} over the radiation field shown in Figure~\ref{f9}
(see \citealt[][equation 2.30]{Osterbrock06}). It is critical to know which part of the 
radiation field dominates the total photoionization rate, since this has the greatest effect on the 
ionization of Component 1. This is shown in the lower panel of Figure~\ref{f11}. 
This panel shows the values of the terms entering in the photoionization rate integral,  namely the product
$\nu^2 \left( 4 \pi J_\nu/h \nu \right) \times \alpha_\nu$. 
In this equation, $\alpha_\nu$ is the opacity and $J_\nu$ is the mean intensity. This shows 
the coupling between the radiation and the gas. Only interactions at energies greater than 13.6 eV affect the ionization of the gas, and the strongest coupling occurs at energies between $\sim200$\,eV to $\sim2$\,keV. When the LOS CF varies, the soft X-rays change, as shown in Figure~\ref{f6}. This carries over into changes in the ionization of He$^0$. This leads to changes in the He$^0$ EUV recombination radiation, which produces the highly ionized species seen by 
{\em HST}.  

The upper panel of Figure~\ref{f11} shows the incident SED as a solid blue line. The solid red line shows the total radiation field, including both the diffuse and attenuated incident, at the midplane of Component 1. This is the net transmitted continuum which is the 5th column of the \texttt{save continuum} command in \texttt{cloudy}. Similar to Figure~\ref{f6}, the effects of diffuse field within the obscurer are included when making a table of the SED passing through the obscurer. We then used this table to generate the appropriate SED in midplane of the Component 1.
The EUV and XUV portions of the SED are heavily extinguished so that most radiation at the midplane is due to diffuse gas emission (Figure~\ref{f9} showed only the diffuse emission). 

As it is obvious from Figure 1,Component 1 is between the Earth and the obscurer so the obscured \cite{Mehd15} SED is what strikes Component 1. The original continuum used here (Figure 3 and Figure 4 red curve) is now available in Cloudy version 17. This observed continuum is quite a bit harder than
the SED in figure 1 of \cite{Math1987}.  The latter was based on observations of very luminous Palomar/Green quasars.   The SED adopted here is based on observations of NGC 5548 obtained in 2013 and 2014 by 
the Anatomy and STORM campaigns. It is worth mentioning that tests show that the choice of intrinsic SED has very little effect on our predictions, which mainly depend on the properties of the obscurer. 

To summarize, we have investigated, in detail, how changes in the LOS CF affect the ionization of the higher-ionization species observed by {\em HST} and identified a unique physical cycle. The LOS CF changes the soft X-ray part of the SED but not the FUV continuum, so the resulting changes would not correlate with the FUV. The soft X-rays change the ionization of helium. The ionizing radiation emitted by recombining He$^+$ changes the ionization rate and abundance of the decorrelated species. However, anything that changes the soft X-rays without affecting the FUV could have a similar effect. This might include the Comptonization scenario outlined by \cite{Mathur17}, the Falling Corona Model studied by \cite{sun18} the partial dust obscuration model of \cite{Gas18}, or other models
such as non-axisymmetic continuum \citep{Dex11} or anisotropic continuum models \citep{Gas03}.

\section{TESTING THE COVERING FACTOR MODEL}
In this paper, we did not try to model any particular observation but examined how changes in the obscurer can affect parts of the SED and result in the observed correlated/decorrelated behavior. We have identified a physical cycle which can reproduce the observed behavior. Here we outline two observational tests of this model. 

\subsection{ Existing observations: The X-ray hardness ratio and inferred LOS CF }
\begin{figure*}
\centering
\includegraphics [width=7in]{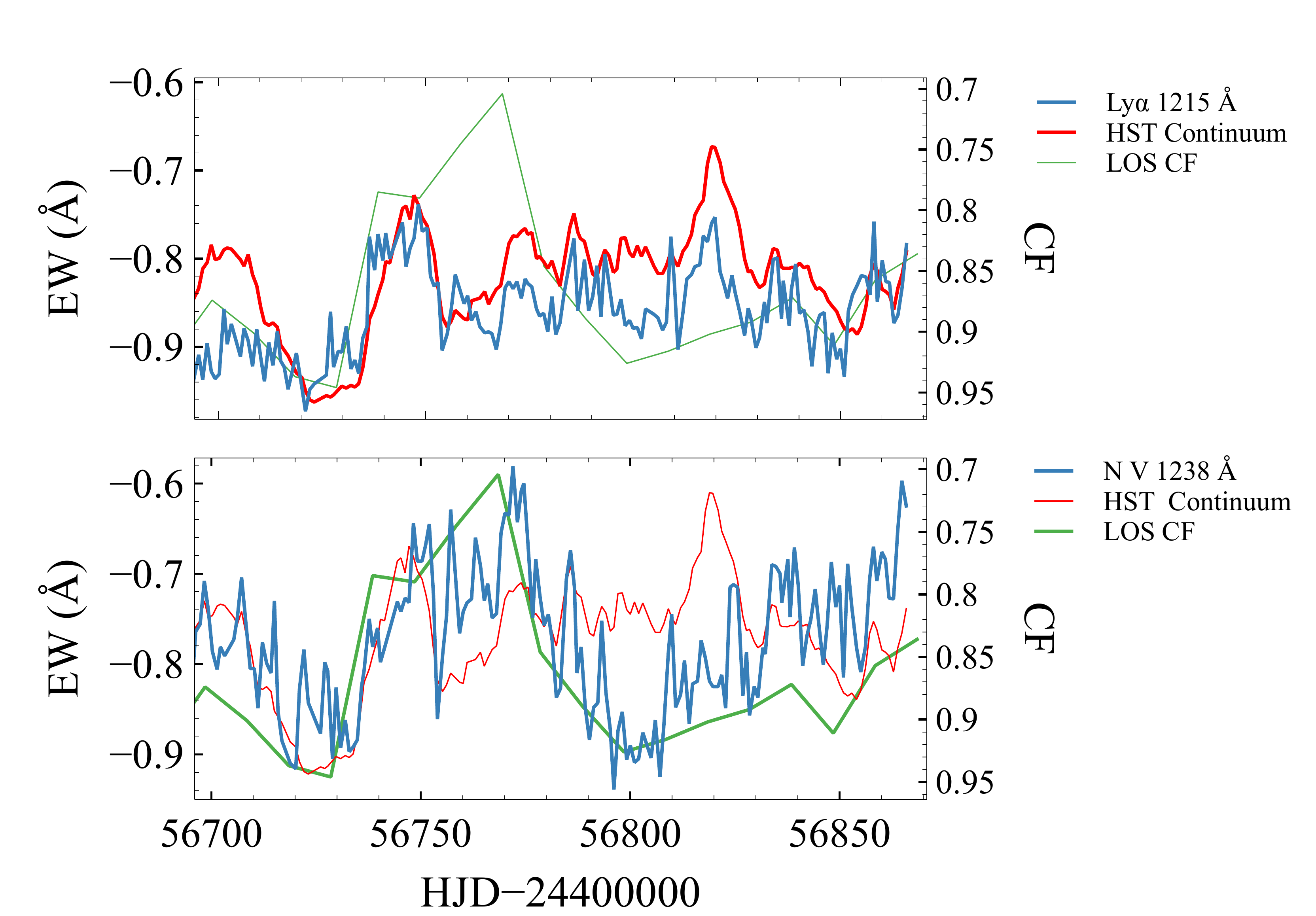}
\caption{The EW of the \lya\ absorption line in Component 1 is shown as the blue line in the upper panel and the 
EW of the corresponding \nv\ absorption line is shown in the lower panel.  The red line shows the {\em HST} FUV 
continuum while the green line is the LOS covering factor of the 
obscurer derived from the X-ray hardness ratio as defined by  \cite{Mehd16}. \lya\ absorption line is correlated with the {\em HST} continuum while 
\nv\ absorption line anticorrelates with the covering factor. }\label{f12}
\end{figure*}

Figure~\ref{f12} summarizes {\em Swift} and {\em HST} observations described by \cite{Kriss18}. The red line is the 
{\em HST} continuum at 1367\,\AA, the blue line shows Ly$\alpha$ absorption line and \nv\ absorption line in the upper and lower panels, respectively. 
These are examples of correlated and decorrelated lines. These are similar to the blue line in the panels of Figure~\ref{f2}. 
In our model, the changing obscurer covering factor is responsible for the absorption-line holiday. The X-ray hardness ratio measured by {\em Swift}, a measure of the hard to soft X-ray brightness, is also a measure of the LOS covering factor, as demonstrated by \cite{Mehd16}, equation 2. The obscurer LOS CF changes, derived by \cite{Mehd16} using the broadband spectral modeling of the {\em Swift} data, are shown in Figure~\ref{f12} as a green line for comparison. The right CF axis is inverted, decreasing from bottom to top, to make it easier to compare with the other quantities plotted. 
The upper panel shows that \lya\ absorption line is correlated with the {\em HST} continuum. The LOS CF is also shown in that panel but with a thinner line to not divert attention. The lower panel is drawn similarly, showing that \nv\  absorption line has a better anti-correlation with the LOS CF rather than the correlation with the 
{\em HST} continuum. 

Figure~\ref{f12} shows that \nv\ absorption responds to variations of the LOS CF better than the {\em HST} continuum. Figure~\ref{f7} shows that larger covering factors and greater extinction cause \nv\ absorption to weaken: \nv\ absorption line is predicted to be anticorrelated with the covering factor. These trends are in the same sense as our predictions.  

\subsection{Future observations:  the full range of obscurer covering factor}
 
\begin{figure*}
\centering
\includegraphics [width=6in]{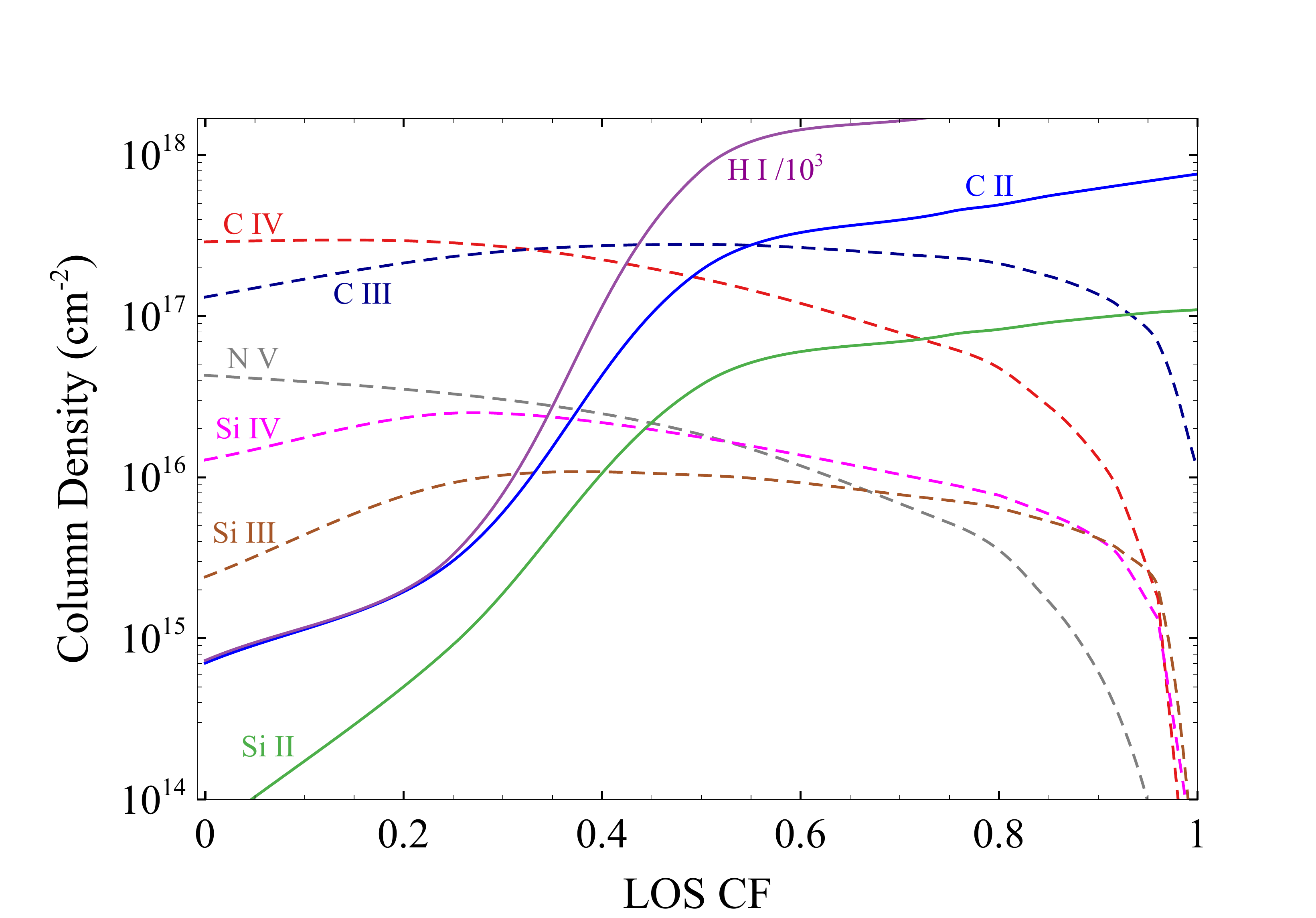}
\caption{The effects of changes in the LOS CF on the ionic column densities over  the full range of covering factor are shown. The dashed and solid lines indicate decorrelated and correlated species, respectively. Their behavior swaps around in the low and high covering factor limits, offering a test of the variable covering factor model. In this models, the ionization parameter for Component 1 is $\log U=-1.5$. }\label{f13}
\end{figure*}

Figure~\ref{f7} focuses on large values of the LOS CF because the covering factor was in this range during the holiday
\citep{Mehd16}. There were other times when the obscurer was not present. Although this was not observed, there must have been times when the obscurer was first coming into our line of sight, and the LOS CF was increasing from small values.  
Figure~\ref{f13} illustrates the full range of the covering factor. The behaviors of the correlated and decorrelated species are reversed for values of LOS~CF in the range 0.3-0.5: The correlated species show dramatic changes while the decorrelated ones remain almost constant. Very small values of the LOS CF represented times before 2011 when there was no obscurer. As Figure~\ref{f13} shows, lower ionization potential species almost disappear. Observations that were performed before 2011 confirm the predictions of Figure~\ref{f13} \citep{Cren09}. 
This motivates future observational tests. Continued monitoring of NGC 5548 by {\em Swift} could identify times when the LOS CF becomes small again. 
{\em HST} observations could then be obtained to follow changes in the absorption lines. 

\vspace{5mm}
\section{SUMMARY }
The reverberation mapping method relies on a causal connection between variations in the lines and continuum. This correlation broke down during the so-called 'holiday' period as discovered by the AGN STORM project. The complications due to these abnormalities may have an effect on derived BLR radii and BH masses, which is why it is important to identify the physics which allows such holidays to occur. The fact that high-ionization absorption lines displayed the holiday while low-ionization absorption lines did not is an important clue to what is happening. It is worth emphasizing that "holiday" was first seen in the Broad emission-lines which have a more complicated geometry \citep{Goad16}.
We showed that changes in the luminosity of the AGN do not produce the observed behavior.  This suggests that changes in the shape of the SED are responsible. 
Strong soft X-ray absorption, produced by a transient cloud referred to as the obscurer, was present throughout the AGN STORM campaign. The obscurer covered only a fraction of the continuum source, which we refer to as the ``line of sight covering factor,'' LOS CF. The soft X-ray absorption was not present before 2011, showing that the LOS CF can change dramatically.  We investigated the effect of a changing SED on  Component 1 cloud producing the strong absorption lines. We have shown that changes in the LOS CF reproduce the observed behavior for large values of the LOS CF. We identified a unique physical cycle in which changes in the LOS CF have a significant effect on soft X-ray portion of the SED. This changes the ionization stage of helium and the ionizing radiation produced as helium recombines drives the changes in the decorrelated absorption lines. Changes in the LOS CF do not affect the optical or UV continuum since the obscurer is transparent that these energies. 
We identified two tests of this model. The first is the {\em Swift} measurements of the X-ray hardness ratio. This can be converted into an obscurer covering factor. This LOS covering factor does seem to correlate with the high ionization ``decorrelated'' absorption lines. We show that the sense of the correlation/decorrelation reverses for smaller covering fractions in the range 0.3-0.5, which can be used to test this scenario in future observations. The tests would have to take place when the covering  factor is very low.
The photoionization models we produced used a variable covering factor to change the soft 
X-ray portion of the SED. However, other models in which the soft X-ray part of the SED changes independently of the optical / UV continuum could produce similar effects. The Comptonization model proposed by \cite{Mathur17} and the Falling Corona Model of \cite{sun18}   could also produce the required changes in the SED. This will be the subject of future work.

\acknowledgments
Support for {\it HST} program number GO-13330 was provided by NASA through a grant from the Space Telescope Science Institute, which is operated by the Association of Universities for Research in Astronomy, Inc., under NASA contract NAS5-26555.We thank NSF (1816537), NASA (ATP 17-0141), and STScI (HST-AR.13914, HST-AR-15018) for their support and Huffaker scholarship for funding the trip to Atlanta to attend the annual AGN STORM meeting, 2017. MC acknowledges support from NASA through STScI grant HST-AR-14556.001-A and STScI grant HST-AR-14286. 
M.D.\ and G.F.\ and F. G.\  acknowledge support from the NSF (AST-1816537), NASA (ATP 17-0141),
and STScI (HST-AR-13914, HST-AR-15018), and the Huffaker Scholarship.
B.M.P., G.D.R., M.M.F, C.J.G., and R.W.P.\ are grateful for the support of the
National Science Foundation through grant AST-1008882 to
The Ohio State University.
A.J.B.\  has  been supported by NSF grant AST-1412693.
M.C.B.\ gratefully acknowledges support through NSF CAREER grant AST-1253702
to Georgia State University.
S.B.\ is supported by NASA through the Chandra award no.\ AR7-18013X issued by
the Chandra X-ray Observatory Center, operated by the Smithsonian
Astrophysical Observatory for and on behalf of NASA under contract
NAS8-03060. S.B.\ was also partially supported by grant HST-AR-13240.009.
E.D.B.\  acknowledge Ssupport from Padua University 
through grants DOR1699945/16, DOR1715817/17, DOR1885254/18, and BIRD164402/16.
K.D.D.\ is supported by an NSF Fellowship awarded under grant AST-1302093.
R.E.\ gratefully acknowledges support from NASA under the ADAP award
80NSSC17K0126.
K.H.\ acknowledges support from STFC grant ST/R000824/1.
SRON is financially supported by NWO, the
Netherlands Organization for Scientific Research.
C.S.K.\  acknowledges the support of NSF grant AST-1009756.
A.P.\ is supported by NASA through Einstein Postdoctoral Fellowship grant number PF5-160141
awarded by the Chandra X-ray Center, which is operated by the Smithsonian Astrophysical Observatory
for NASA under contract NAS8-03060.
T.T.\ has been supported by NSF grant AST-1412315.
T.T.\ and B.C.K.\ acknowledge support from the Packard Foundation in the form of
a Packard Research Fellowship to T.T.
The American Academy in Rome and the
Observatory of Monteporzio Catone are thanked by T.T.\ for kind hospitality. 
M.V.\ gratefully acknowledges support from the
Danish Council for Independent Research via grant no.\ DFF 4002-00275.
This research has made use of the NASA/IPAC Extragalactic Database (NED),
which is operated by the Jet Propulsion Laboratory, California Institute of Technology, under contract with the National Aeronautics and Space Administration. I. McH.\ acknowledges support from a Royal Society Leverhulme Trust Senior Research Fellowship LT160006 and from STFC grant ST/M001326/1. J.M.G. gratefully acknowledges support from NASA under awards NNX15AH49G and 80NSSC17K0126. E.D.B. acknowledges support from Padua University through grants DOR1699945/16, DOR1715817/17, and DOR1885254/18.

 \end{document}